\documentclass[%
 reprint,
 amsmath,amssymb,
 aps,
]{revtex4-1}

\usepackage{graphicx}
\usepackage{dcolumn}
\usepackage{bm}

\begin{document}

\preprint{APS/123-QED}

\title{Non-covalent quantum machine learning corrections to density functionals}

\author{P\'{a}l D. Mezei}
\author{O. Anatole von Lilienfeld}
\email{anatole.vonlilienfeld@unibas.ch}
\affiliation{Institute of Physical Chemistry and National Center for Computational Design and Discovery of Novel Materials,
Department of Chemistry, University of Basel, Basel, Switzerland}
\date{\today}

\begin{abstract}
We present non-covalent quantum machine learning corrections to six physically motivated density functionals with systematic errors.
We demonstrate that the missing massively non-local and non-additive physical effects can be recovered by the quantum machine learning models.
The models seamlessly account for various types of non-covalent interactions, and enable accurate predictions of dissociation curves.
The correction improves the description of molecular two- and three-body interactions crucial in large water clusters, and provides a reasonable atomic-resolution picture about the interaction energy errors of approximate density functionals that can be a useful information in the development of more accurate density functionals.
We show that given sufficient training instances the correction is more flexible than standard molecular mechanical dispersion corrections, and thus it can be applied for cases where many dispersion corrected density functionals fail, such as hydrogen bonding.
\end{abstract}

\maketitle

The efficient and accurate calculation of non-covalent interactions (exchange repulsion, electrostatics, induction, and dispersion \cite{jeziorski1994}) is crucial in many areas of physics such as in the description of surface adsorptions and reactions, layered structures, organic molecular crystals, polymer crystals, soft matter, diffusive motions, and the dynamic properties of water.
However, the highly non-local nature of these interactions makes them difficult to capture by efficient computational methods such as many density functional (DF) approximations.\cite{kristyan1994,hobza1995,perez1995}
The leading terms in the long-range dispersion interaction energy can simply be described by the two-body London \cite{london1937,casimir1948} and three-body Axilrod-Teller \cite{axilrod1943,aubzienau} potentials, and hence the missing dispersion interaction is often recovered by molecular mechanical dispersion corrections (\textit{e.g.}, TS \cite{ts09,lilienfeld2010} or D3 \cite{dftd3,dftd3bj}), or by nonlocal van der Waals DFs.\cite{vdwdf,vdwdf2,vv09,vv10}
However, it is especially difficult to correct semi-local functionals simultaneously suffering from delocalization and dispersion errors at non-covalent overlapping electron density regions.\cite{dfterrors,darctest,anionpitest}
In such cases, one may consider more expensive fifth-rung functionals (on the Jacob's ladder of DF approximations \cite{jacob}) also containing information about the virtual orbitals.\cite{b2plyp,dsdpbehb95,xyg3,pwrb95,drpa75,scsdrpa75,janesko2009}

An alternative approach is to use efficient quantum machine learning (QML) models with representations uniquely describing the electronic structure problem to solve (for a given number of electrons without external fields).\cite{qml,coulmat,chemspace}
Previously, such models were only applied for correcting approximate DF one- and two-body total energy errors on a set of water clusters and ice structures, or relative conformational enthalpy errors on a set of C$_{7}$H$_{10}$O$_{2}$ isomers;\cite{gillan2013,ramakrishnan2015} however, no universal QML model has been presented yet for correcting simultaneously the inter- and intramolecular interaction energies of approximate DFs.
Here, we propose non-covalent interaction (NCI) corrections from QML with high level of flexibility to correct the approximate DF description of intermolecular interactions, going beyond pure dispersion.
We demonstrate that the missing physical effects can be learned by these NCI models.
We analyze the errors of the resulting models from multiple aspects, and compare them to similarly efficient dispersion corrected functionals.
Note that other machine learning (ML) models were also applied to improve upon approximate quantum chemical non-covalent interactions,\cite{gao2016,mp2nn} but they generally lack uniqueness in the representation and thus can yield absurd results.\cite{fourier}

In this paper, we use KRR (in the QML code \cite{qmlcode}) with Gaussian kernel and the recently introduced representation of Faber et al. (FCHL18).\cite{fchl}
We perform a two-step procedure to decompose the correction into atomic contributions.
In the first step, we determined the regression coefficient vector \text{\boldmath{$\alpha$}},
\begin{equation}\label{step1} \text{\boldmath{$\alpha$}} = (\textbf{K} + \lambda \textbf{I})^{-1} \textbf{y} \end{equation}
where \textbf{K} is a local symmetric kernel matrix with the matrix elements of $\sum\limits_{I\epsilon i} \sum\limits_{J\epsilon j} k(x^{i}_{I},x^{j}_{J})$ using the \textit{I}th atomic elements of the \textit{i}th training systems (molecular dimers and monomers) from the \textbf{x} array of representation vectors, $\lambda$ is the regularization strength (chosen to be $\lambda$ = $10^{-8}$ due to the very low level of noise in the well-converged CCSD(T) reference energies; for more information see the supplemental material), \textbf{I} is the identity matrix, and \textbf{y} is the training output vector.
In the second step, we predict atomic contributions $\textbf{$\tilde{y}$}^{i}_{I}$ for the \textit{I}th atom of the \textit{i}th test system (as well as their sum),
\begin{equation}\label{step2} \textbf{$\tilde{y}$}^{i}_{I} = \sum\limits_{j} {\alpha}_j \sum\limits_{J\epsilon j} k(\tilde{x}^{i}_{I},x^{j}_{J}) \end{equation}
where the atomic kernel matrix elements $k(\tilde{x}^{i}_{I},x^{j}_{J})=\exp(-\frac{||\tilde{x}^{i}_{I} - x^{j}_{J}||_2^2}{2\sigma^2})$ (with an optimal hyperparameter $\sigma$ = $2^{8}$) are calculated from the training \textbf{x} and test $\mathbf{\tilde{x}}$ arrays of atomic representation vectors.
Note that a similar but atom-pairwise decomposition was already presented with the bag of bonds representation to extract interatomic potentials from a trained machine learning model.\cite{bob}

As baseline methods, we selected from each of the second-, third-, and fourth-rung a usually more repulsive (BLYP,\cite{b88,lyp} TPSS,\cite{tpss} B3LYP\cite{b3lyp}) and a usually less repulsive (PBE,\cite{pbe} SCAN,\cite{scan,scan2} PBE0\cite{pbe0}) functional with physically motivated forms and thus expectedly with systematic errors.
We also expect better description of the midrange correlation from the less repulsive functionals, and reduced self-interaction error from functionals on higher rungs.
In order to be able to generalize our correction to also improve intramolecular interactions, we trained the machine learning model on atomization energies, and derived the interaction energy correction from the atomization energy corrections for monomers and dimer/cluster.

\begin{figure}[h]\includegraphics[width=\linewidth]{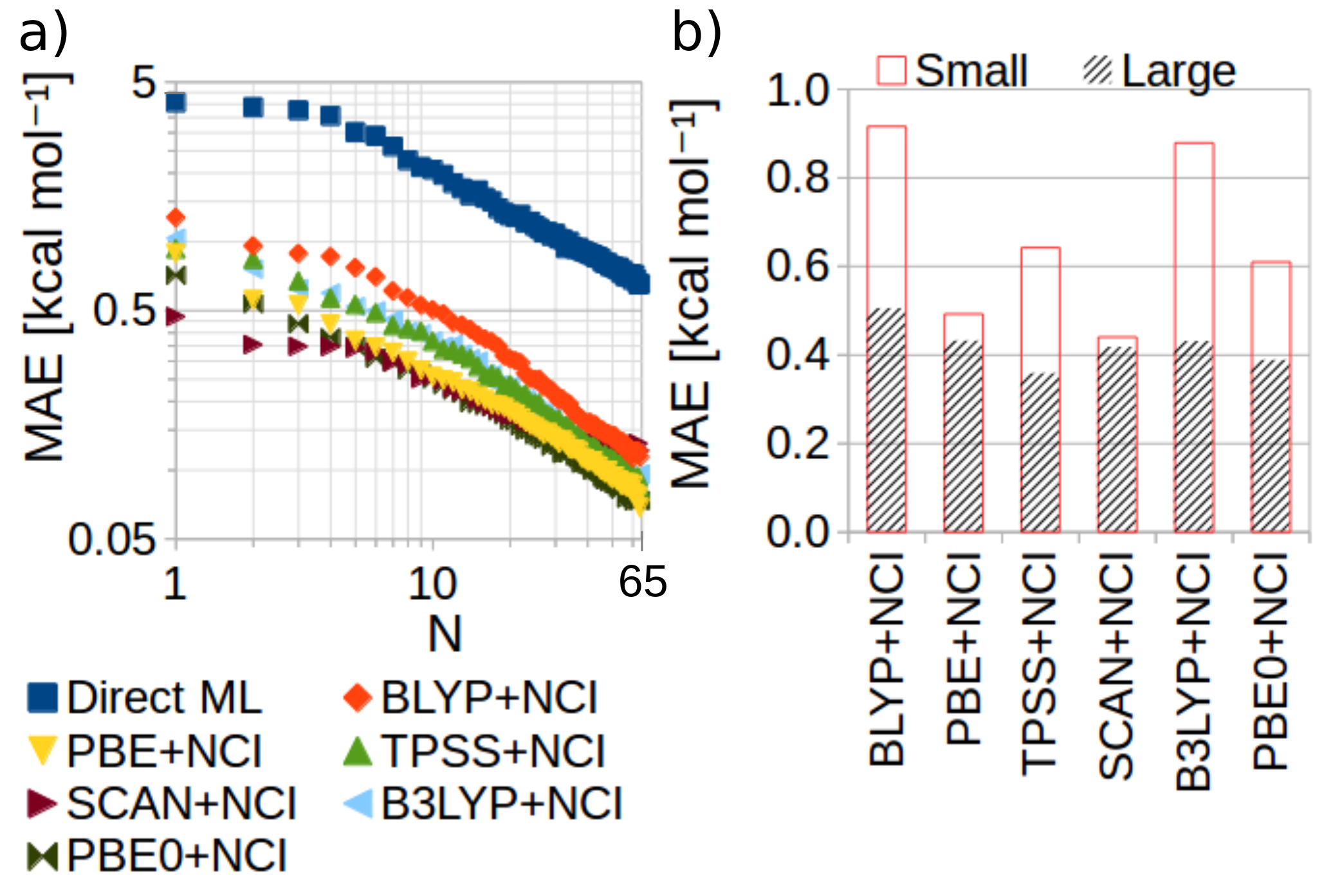}\caption{\label{learning}a) Learning curves (from 100 random cross validation runs adequate for varying training/test proportion) for the direct ML and NCI models on S66x8 using N and (66-N) potential curves in training and test set, respectively. b) Overall mean absolute errors (weighted by the number of equilibrium dimers/clusters in the datasets) on a broader dataset training the NCI models only on S66x8 (small) and on the broader dataset (large).}\end{figure}

First, we present learning curves (FIG. \ref{learning}a) for training the NCI models on the biochemically relevant S66x8 set.
As expected,\cite{ramakrishnan2015} the learning curves of the NCI models run more or less parallel shifted downwards by a factor depending on the inherent accuracy of the baseline methods compared to the learning curve corresponding to the direct ML model.
The learning onset is reached after including 5 potential curves into the training set.
Despite the already promising overall accuracy of the NCI models trained on S66x8, using only S66x8 in the training set leads to less transferable models biased towards small dimers with H, C, N, and O atoms and few functional groups.

To achieve a more transferable model, we extended our training set by further potential curves and water clusters (FIG. \ref{learning}b, for more details see supplementary material).
We will use this extended training set hereafter.
On each rung, among the selected functionals, it is easier to correct the less repulsive one with the NCI model than the corresponding more repulsive one.
From the second and third rungs, the NCI corrected models based on the less repulsive functionals usually outperform the models based on the more repulsive functionals from the third and fourth rungs.
Among the DFs considered, the least robust model is BLYP+NCI, while the most robust one is SCAN+NCI.
We will analyze the latter in the remainder of this paper.

\begingroup
\begin{table}[h]
\caption{\label{dftd3}Mean absolute errors (kcal mol$^{-1}$) in the interaction energies for the different test sets with our most robust NCI approach or D3 corrections.}
\begin{center}
\begin{tabular}{ c c c c }
 \hline
 \bf{Dataset} & \bf{SCAN} & \bf{+NCI} & \bf{+D3} \\
 \hline
 Biochemically relevant \cite{s66,s66x8}  & 0.70 & 0.14 & 0.25 \\
 Non-covalent blind test \cite{bt80,bt80dft} & 0.40 & 0.25 & 0.16 \\
 Halogen-containing \cite{x40,x40x10martin}      & 0.49 & 0.24 & 0.36 \\
 Water clusters \cite{temelso2011,water27,anacker2014,manna2017}         & 4.85 & 0.94 & 6.89 \\
 Molecular crystals$^a$ \cite{x23,x23geom,x23mbd,x23pbed3cbs,x23b3lypdst}      & 4.69 & 2.38 & 2.20 \\
 Host-guest complexes$^a$ \cite{s12l,s12lgeom}   & 9.98 & 3.66 & 1.35 \\
  \hline
\end{tabular}
\end{center}
\raggedright $^a$ Not represented in the training set.
\end{table}
\endgroup

In a first step, we compare the accuracy of the SCAN+NCI interaction energies to the accuracy of the corresponding D3 dispersion corrected interaction energies \cite{dftd3bj} with Becke-Johnson (BJ) damping and three-body terms (TABLE \ref{dftd3}, more details in the supplemental material).
The NCI correction usually works better than the D3 correction for small dimers and even for large water clusters, but it is worse for large host-guest complexes and molecular crystals.
The success of the NCI corrected schemes on water clusters suggests that the NCI models can correct errors also other than dispersion.
For these neutral water clusters, our NCI correction is on par with the most accurate dispersion corrected DF (MAE: 0.94 kcal mol$^{-1}$ with PW6B95-D3(0)) \cite{pw6b95,dftd3,gmtkn30} from the literature.

\begin{figure}[h!]\includegraphics[width=\linewidth]{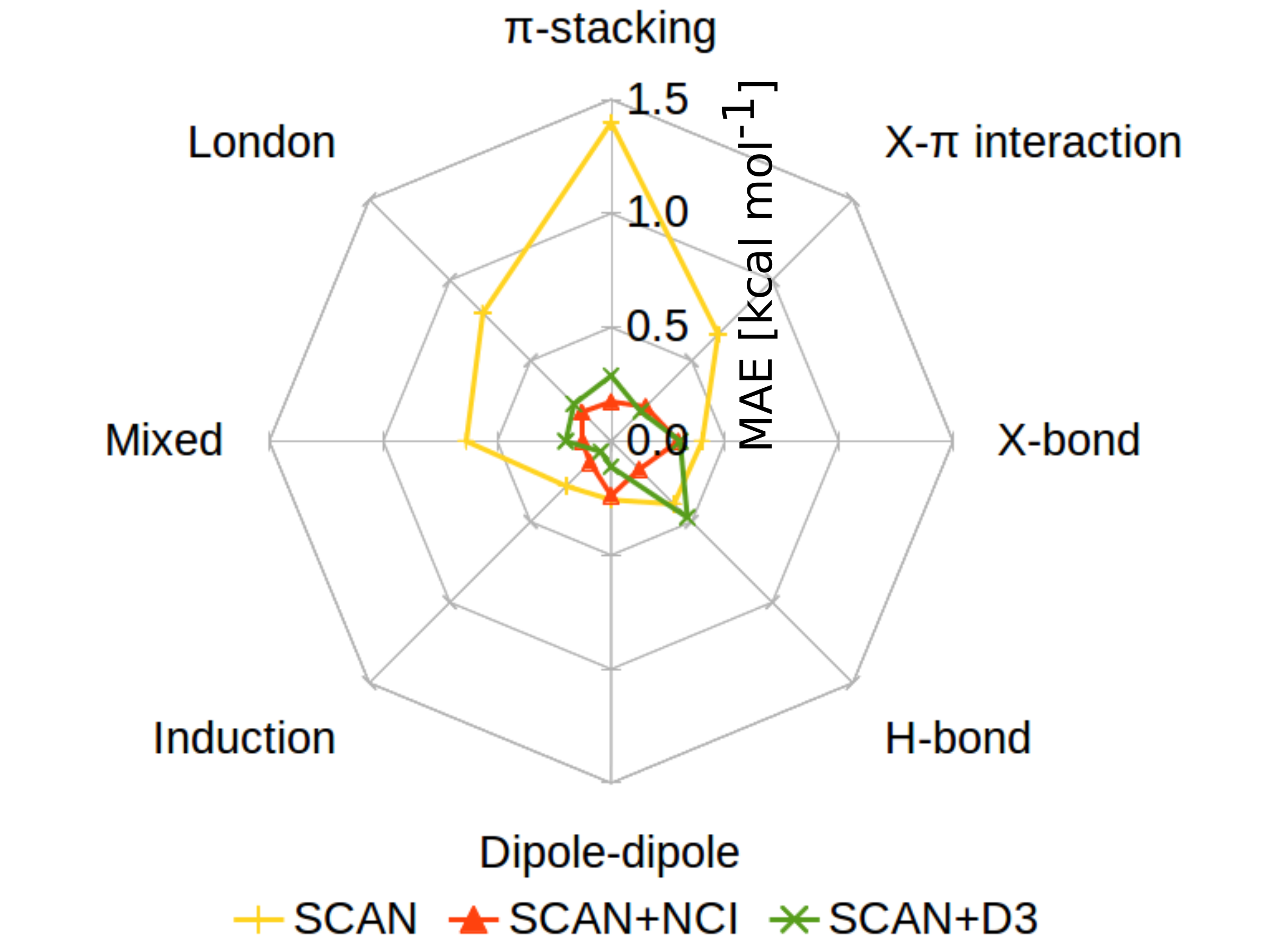}\caption{\label{interaction}Mean absolute errors (kcal mol$^{-1}$) the pure, NCI, and D3 corrected SCAN methods on various types of interactions in biochemically relevant and halogen-containing (X: halogen atom) dimers.\cite{s66x8,x40x10martin}}\end{figure}

The NCI correction performs well for various types of interactions (FIG. \ref{interaction}).
The SCAN+NCI interaction energies are more accurate than the SCAN+D3 ones for hydrogen-bonding and dispersion but less accurate for dipole-dipole interaction and induction.
Note that hydrogen bonds can be recognized by three atoms, while the dipole moment is a global property of molecules that is poorly described by local representations, as already discussed in refs \cite{fchl,faber2017prediction}.
In the other cases, the two correction schemes work similarly.

\begin{figure}[h!]\includegraphics[width=\linewidth]{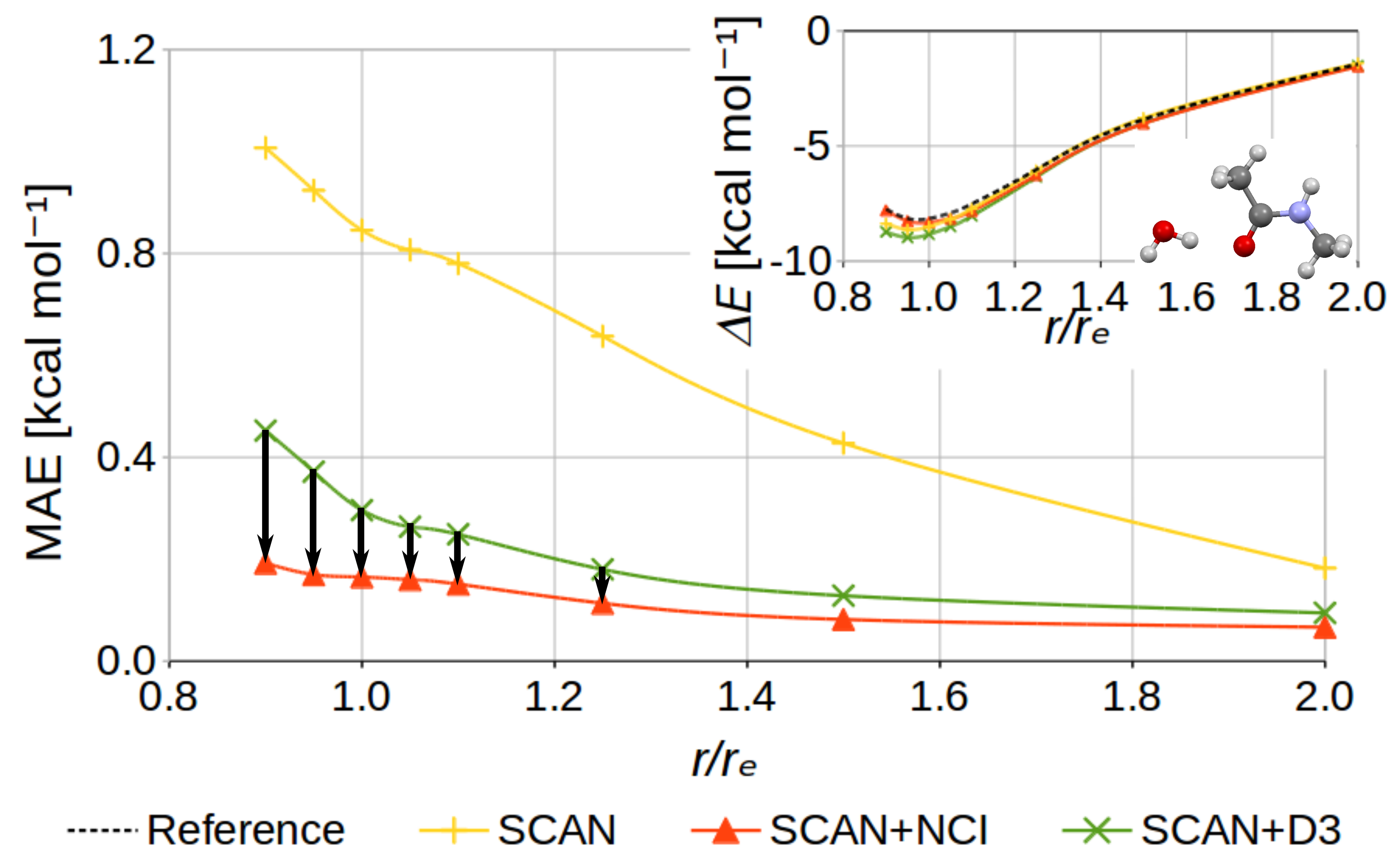}\caption{\label{distance}Mean absolute errors (kcal mol$^{-1}$) with respect to the relative intermolecular distance on the S66x8 database for the pure, NCI, and D3 corrected SCAN methods. The potential energy curves for a water-peptide dimer can be compared to the CCSD(T) reference in the inset. (The arrows represent the physical effects missing from the D3 but described by the NCI correction.)}\end{figure}

In a second step, we have examined how the NCI model works at non-equilibrium intermolecular distances (FIG. \ref{distance}).
The NCI correction works well at various relative intermolecular distances (defined as the actual (\textit{r}) over equilibrium ($\textit{r}_e$) distance between the two monomer centers of mass).
While the D3 correction has difficulties at shorter relative intermolecular distances (especially for hydrogen-bonding), where density functionals often overestimate the magnitude of the interaction because of their delocalization error.

\begin{figure}[h!]\includegraphics[width=\linewidth]{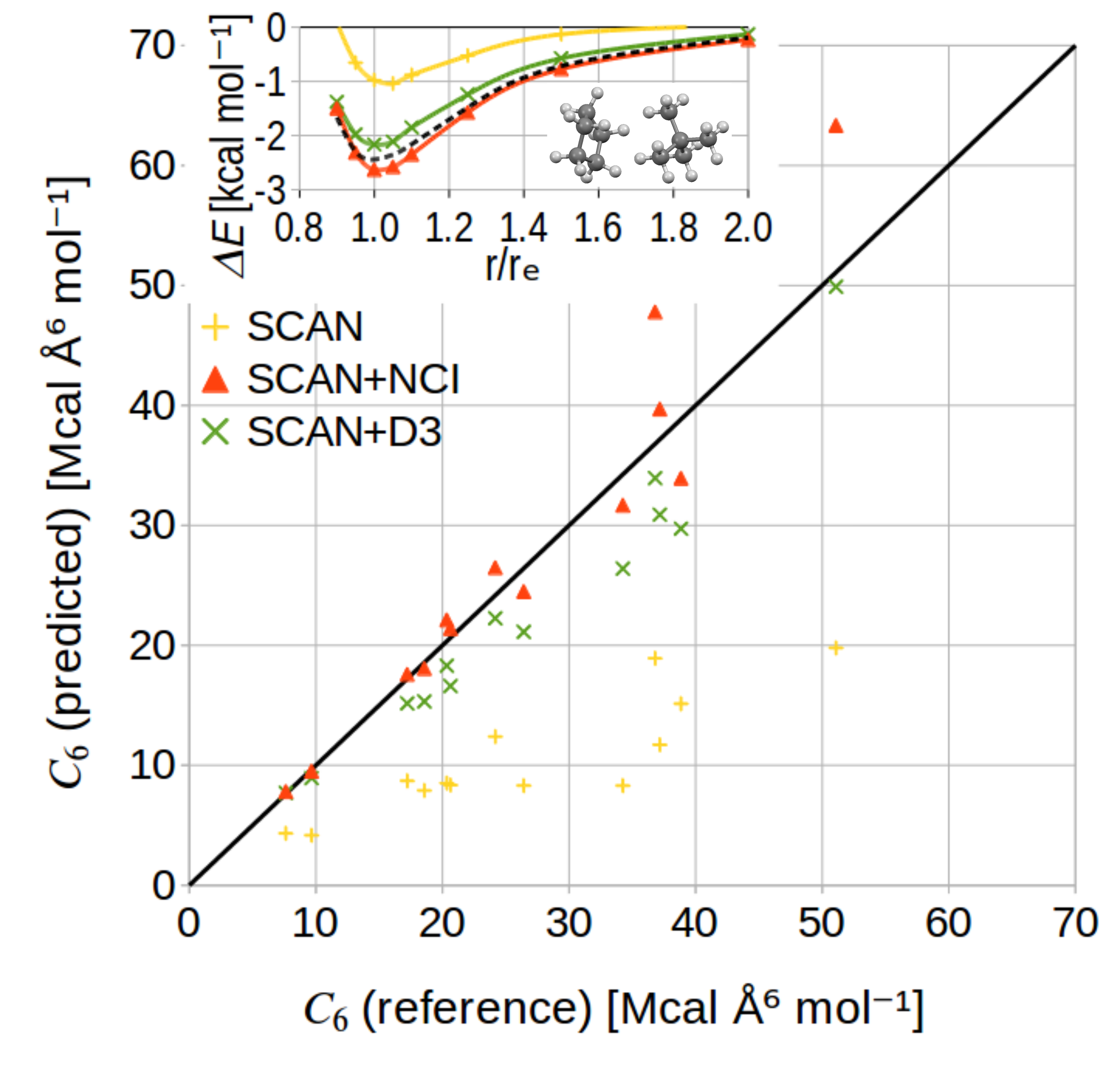}\caption{\label{dispersion}Predicted versus reference seeming molecular two-body $C_6$ dispersion coefficients (kcal mol$^{-1}$) of S66x8 dimers with London dispersion for the pure, NCI, and D3 corrected SCAN methods. The potential energy curves for the cyclopentane-neopentane dimer can be compared to the CCSD(T) reference (black dashed line) in the inset.}\end{figure}

The SCAN+NCI seeming molecular two-body $C_6$ dispersion coefficients (fitting a $-C_6/r^6$ function to the three longest intermolecular distances, the actual dispersion coefficients may be smaller) of the London dispersion dominated S66x8 hetero- and homodimers are in good agreement with the reference (FIG. \ref{dispersion}).
Even though, the NCI correction does not contain explicitly information neither about the long-range behavior of dispersion nor about the values of the dispersion coefficients, it can lead to a similarly accurate and meaningful description of long-range dispersion as the D3 correction.

\begin{figure}[h!]\includegraphics[width=\linewidth]{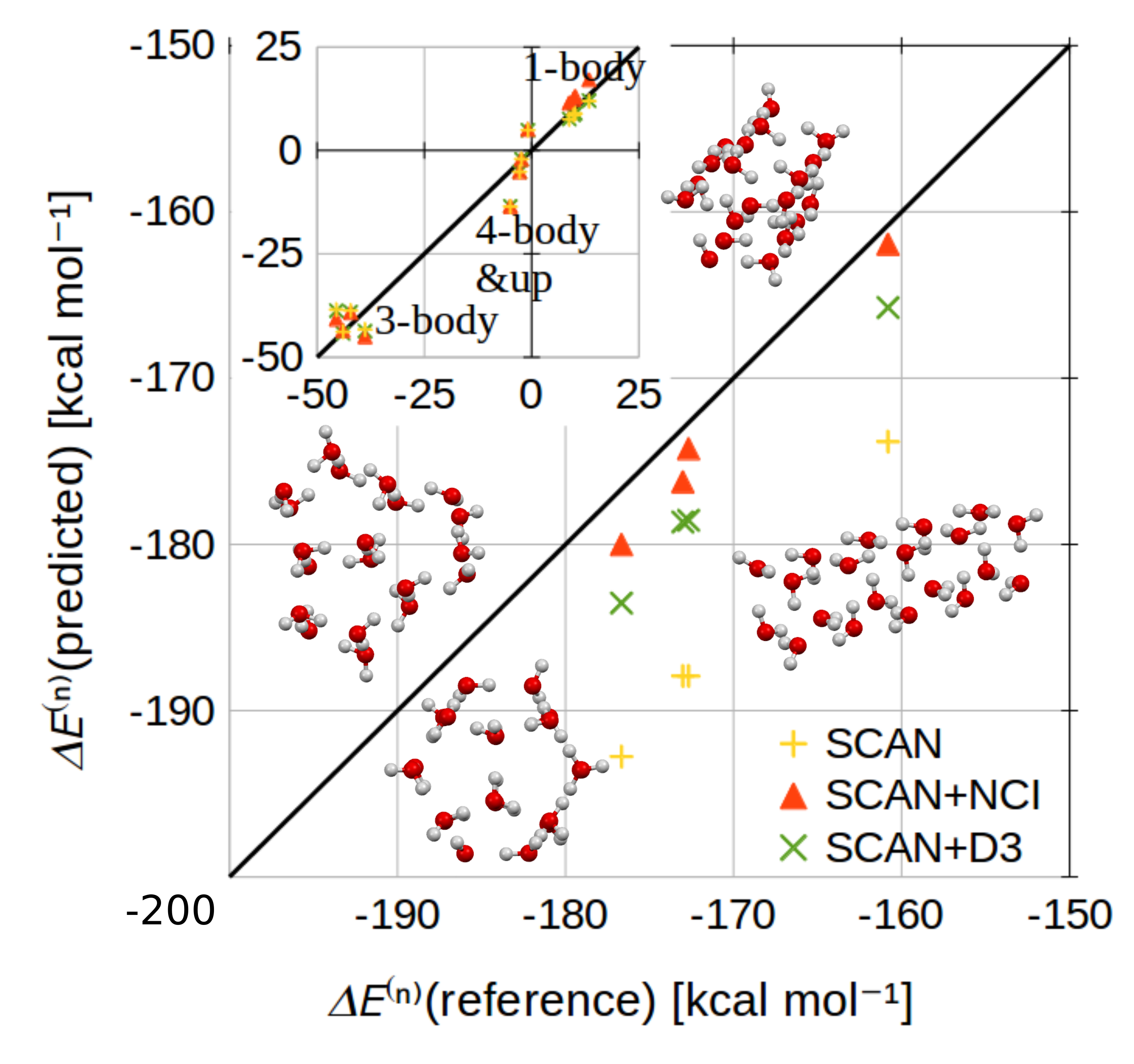}\caption{\label{manybody}Predicted versus reference CCSD(T) molecular two-body interaction energy terms (kcal mol$^{-1}$) of the water icosamers for the pure, NCI, and D3 corrected SCAN methods. To provide context, the molecular one-, three-, and higher many-body interaction energy terms are shown in the inset.}\end{figure}

In a third step, the molecular many-body decomposition of the NCI corrected interaction energies on the water icosamers. The results in FIG. \ref{manybody} show that the largest correction appears in the two-body term.
The NCI model goes beyond the atom-pairwise picture, which provides more flexibility to properly account for the accurate description of the interaction energy in water clusters.
The correction does not affect the four-body and higher-order terms, because the FCHL representation contains information only up to atomic three-body.
Hence the good overall accuracy relies on an error cancellation between the molecular three-body and higher-order terms. 

\begin{figure}[h!]\includegraphics[width=\linewidth]{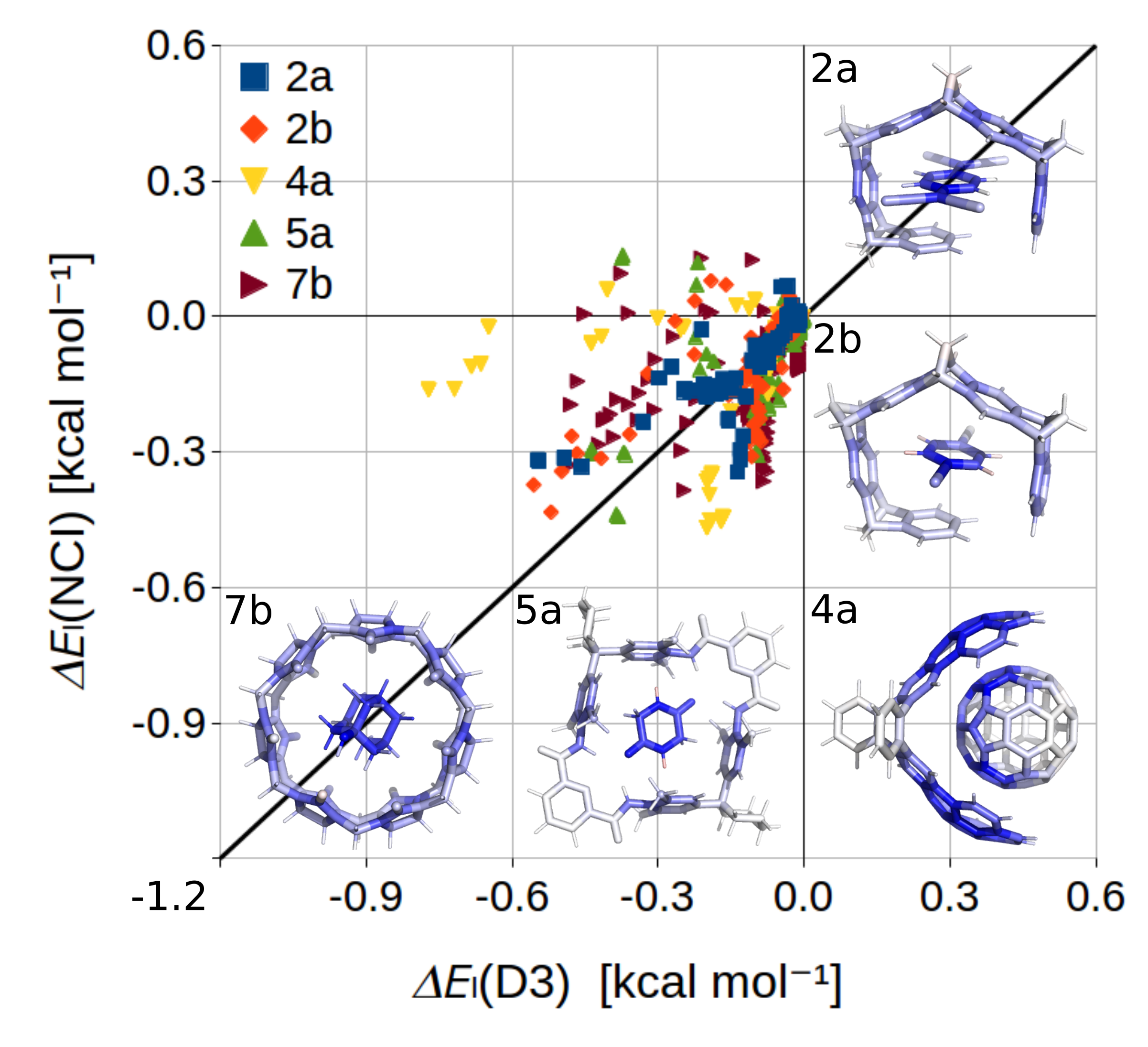}\caption{\label{atomic}Atomic contributions (kcal mol$^{-1}$) of NCI versus D3 corrections of SCAN for five large host-guest complexes.\cite{s12l,s12lgeom} The NCI atomic contributions are mapped on the molecules in inset (blue: attractive; red: repulsive).}\end{figure}

\begin{figure}[h!]\includegraphics[width=\linewidth]{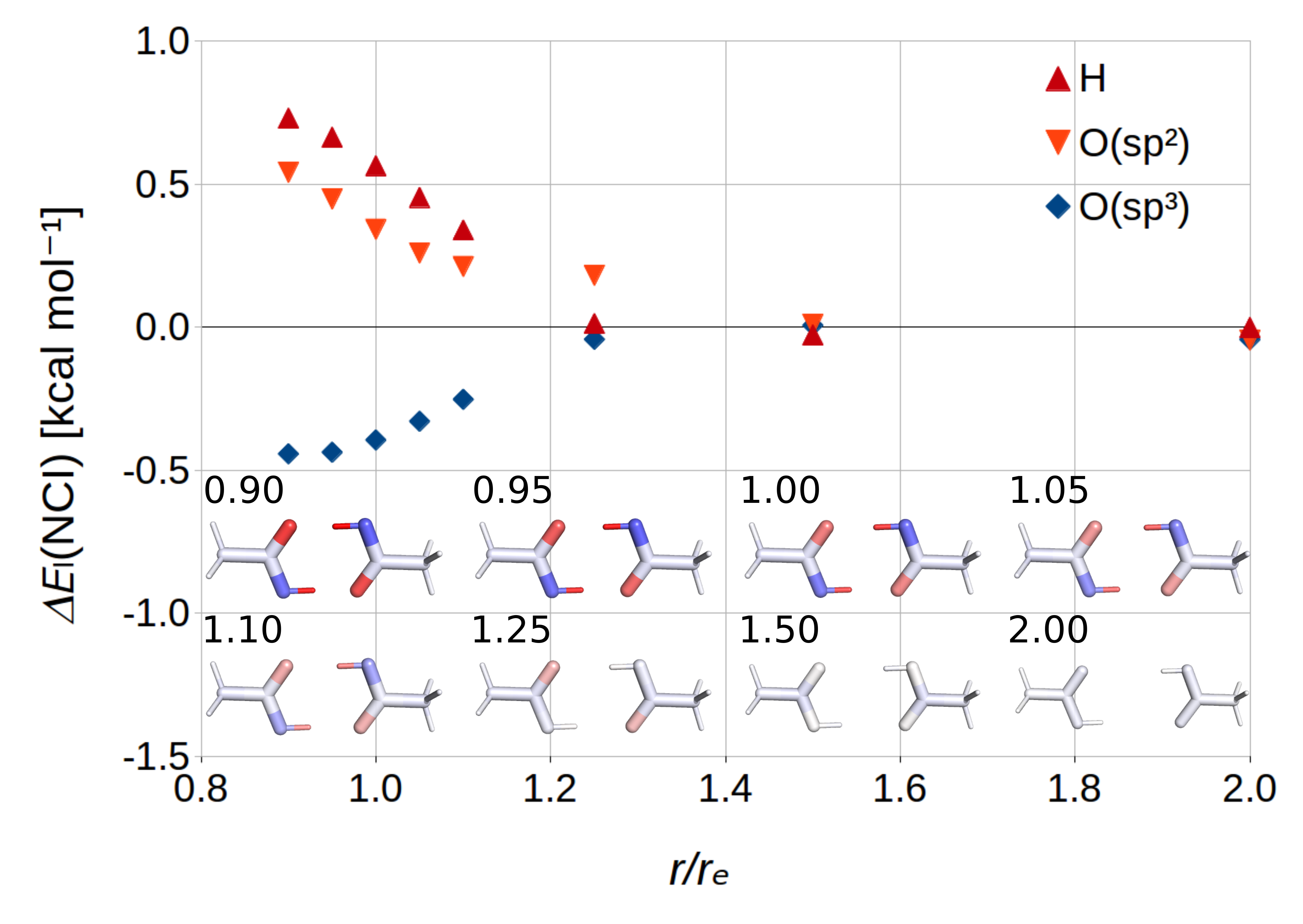}\caption{\label{contributions}Atomic contributions (kcal mol$^{-1}$) of the NCI correction for the acetic acid dimer with respect to the relative intermolecular distance. The NCI atomic contributions are mapped on the dimers in inset (blue: attractive; red: repulsive).}\end{figure}

Finally, we have also extracted atomic level information from the NCI models by exploiting the atomic kernel matrix in the prediction.
The atomic contributions (between an atom in monomer A and all the atoms in monomer B) to the NCI interaction energy correction (without monomer relaxation) in large host-guest complexes (FIG. \ref{atomic}) correlate well with the pairwise atomic contributions to the D3 correction ($\Delta E^{i}_{I}=\frac{1}{2}\sum\limits_{J\in j} E^{IJ}_{6}$) for smaller interaction energy contributions.
The randomity in the scatter plot of the NCI versus D3 atomic contributions is higher, however, for larger interaction energy contributions.
(Note that the atomic contributions to the two corrections do not have to agree necessarily since the atomic decomposition of the interaction energy is ambiguous.
Also note that the host-guest complexes in FIG. \ref{atomic} were not represented in the training set.)
In general, the NCI model places the attractive correction on the more polarizable (non-hydrogen) atoms, or on atoms involved in $\pi$ delocalization, and the repulsive (or less attractive) correction on atoms of different monomers close to each other, or on atoms involved in hydrogen bonding.
The decay of the atomic contributions with respect to the relative intermolecular distance can be followed for a hydrogen-bonded dimer in FIG. \ref{contributions}.

We applied QML to correct the non-covalent interactions in various van der Waals complexes calculated by efficient DF approximations.
Learning curves demonstrate that the NCI corrections can seamlessly complement approximate DFs for various types of interactions and at various relative intermolecular distances.
The correction captures massively non-local and non-additive physical effects missing from the DF methods, has an effect on the first three molecular many-body interaction energy terms, and provides a reasonable atomic-resolution picture about the interaction energy errors of DF approximations.
For these reasons, we think our NCI model can also become a useful tool in the development of efficient DF approximations to understand the nature of the interaction energy errors.
Among the examined NCI models, the most robust is based on the SCAN functional, which can recognize non-covalent electron density overlaps.
The NCI correction is more flexible than standard molecular mechanical dispersion corrections, and it can capture more of the missing physical effects than the generally used D3 correction.
As evinced, the NCI correction is more accurate than conventional dispersion corrections for example when hydrogen bonding dominates the interaction.
Furthermore, its error decreases systematically if more training data are provided thanks to the generalization power of QML models.\cite{qml}
To facilitate applications, we make our machine learning models available in the supplementary material.
Future improvements of the NCI corrections will be possible as soon as more high-level reference non-covalent interaction energies become available.
Another possible extension of this work could be its extension within the recently introduced multi-level combination technique.\cite{zaspel}

PDM thanks Jan Gerit Brandenburg for providing data and information about the calculations on the S66x8 molecular dimers and the X23 molecular crystals.
This research was supported by the NCCR MARVEL, funded by the Swiss National Science Foundation.
This project has received funding from the European Research Council (ERC) under the European Union's Horizon 2020 research and innovation programme under grant agreement No 772834.

\bibliography{refs.bib}

\end{document}